\documentclass[11pt]{article}
\usepackage{jcappub,natbib,float,caption,comment,bm}

\def\be{\begin{equation}}
\def\ee{\end{equation}}
\def\ba#1\ea{\begin{align}#1\end{align}}
\newcommand{\vs}{\nonumber\\}

\renewcommand{\v}[1]{\mathbf{#1}}
\newcommand{\vx}{\v{x}}
\newcommand{\vr}{\v{r}}
\newcommand{\vk}{\v{k}}
\newcommand{\vq}{\v{q}}

\newcommand{\refeq}[1]{eq.~(\ref{eq:#1})}          
          
\newcommand{\refEq}[1]{Eq.~(\ref{eq:#1})}          
\newcommand{\reffig}[1]{figure~\ref{fig:#1}}          
\newcommand{\refFig}[1]{Figure~\ref{fig:#1}}          
\newcommand{\refsec}[1]{section~\ref{sec:#1}}          
\newcommand{\refapp}[1]{appendix~\ref{app:#1}}

\newcommand{\Om}{\Omega_m}

\newcommand{\rhob}{\bar\rho}

\renewcommand{\d}{\delta}
\newcommand{\tOm}{\tilde{\Omega}_m}
\newcommand{\tOK}{\tilde{\Omega}_K}

\newcommand{\cO}{\mathcal O}

\title{Position-dependent power spectrum of the large-scale structure:
a novel method to measure the squeezed-limit bispectrum}
\author[a]{Chi-Ting Chiang,}
\author[a]{Christian Wagner,}
\author[a]{Fabian Schmidt,}
\author[a,b]{Eiichiro Komatsu}
\affiliation[a]{Max-Planck-Institut f\"ur Astrophysik, Karl-Schwarzschild-Str. 1, 85741 Garching, Germany}
\affiliation[b]{Kavli Institute for the Physics and Mathematics of the
Universe, Todai Institutes for Advanced Study, the University of Tokyo,
Kashiwa, Japan 277-8583 (Kavli IPMU, WPI)}
\emailAdd{ctchiang@mpa-garching.mpg.de}
\abstract{
The influence of large-scale density fluctuations on structure formation
on small scales is described by the three-point correlation function
(bispectrum) in the so-called ``squeezed configurations,'' in which one
wavenumber, say $k_3$, is much smaller than the other two, i.e., $k_3\ll
k_1\approx k_2$. This bispectrum is generated by non-linear
gravitational evolution and possibly also by inflationary 
physics. In this paper, we use this fact to show that the bispectrum in
the squeezed configurations can be measured {\it without} 
employing three-point function estimators.
Specifically, we use the ``position-dependent
power spectrum,'' i.e., the power spectrum measured in smaller
subvolumes of the survey (or simulation box), and correlate it with the
mean overdensity of the corresponding subvolume. This correlation
directly measures an integral of the bispectrum dominated by
the squeezed configurations. Measuring this correlation is only slightly
more complex than measuring the power spectrum itself, and sidesteps the
considerable complexity of the full bispectrum estimation. We use
cosmological $N$-body simulations of collisionless particles with
Gaussian initial conditions to show that the measured correlation
between the position-dependent power spectrum and the long-wavelength
overdensity agrees with the theoretical expectation. The position-dependent
power spectrum thus provides a new, efficient, and promising way to measure
the squeezed-limit bispectrum from large-scale structure observations such
as galaxy redshift surveys.
}
\begin{document}
\maketitle
\flushbottom

\section{Introduction}

Suppose that we measure a two-point correlation function (power
spectrum) of density fluctuations in the Universe. We normally measure
this quantity from the entire survey volume in which we have
measurements of the matter distribution. Let us divide the survey volume
into many subvolumes and measure the power spectrum in each subvolume. In
this paper, we show that the power spectrum in each subvolume depends
on environment, and is specifically correlated
with the mean overdensity of that subvolume.
This correlation measures how the small-scale power spectrum responds to the 
presence of a large-scale density fluctuation, which can be equivalently
described by a non-vanishing three-point function (bispectrum).

Even if the initial density fluctuations generated by inflation are perfectly
Gaussian, the subsequent non-linear gravitational evolution
of matter generates a non-zero bispectrum (see
\cite{bernardeau/etal:2001} for a review). The ``position-dependent
power spectrum'' thus offers a test of our understanding of structure
formation in the Universe. Moreover, improving our understanding of
structure formation increases the sensitivity to a small bispectrum
generated by inflation, making it possible to test the physics of
inflation using observations of the large-scale structure of the Universe.

Not only is this new observable of the large-scale structure of the
Universe conceptually straightforward to interpret, but it is also 
simpler to measure than the full bispectrum. Constraining the physics of
inflation using the squeezed-limit bispectrum of the cosmic microwave
background is a solved problem \cite{komatsu:2010}. However, doing the
same using the bispectrum of the large-scale structure (e.g.,
distribution of galaxies) is considerably more challenging due to
complex survey selection functions as well as to mode couplings caused
by the non-linearity of the matter density field as well as the 
complexity of galaxy formation.  This explains why only
few measurements of the bispectrum have been reported in the literature
\cite{scoccimarro/etal:2000,verde/etal:2001,nishimichi/etal:2006}, 
and further motivates our use of the position-dependent power spectrum as a
simpler route to measuring the squeezed-limit bispectrum. While we
mostly have galaxy redshift surveys in mind, this idea can also be applied
to the projected matter density as measured through lensing.

In this paper, we study this new observable. We show that the
position-dependent power spectrum measures an integral of the
bispectrum, which is dominated by the bispectrum in the so-called
``squeezed configurations,'' in which one wavenumber, say $k_3$, is much
smaller than the other two, i.e., $k_3\ll k_1\approx k_2$. This limit of
the bispectrum has a straightforward interpretation (i.e., the
large-scale density fluctuation modulating the small-scale power
spectrum), which can be predicted using a simple calculation. We
restrict ourselves to the position-dependent power spectrum of
collisionless particles in real space in this paper.  We shall
incorporate the effects of halo bias and redshift-space distortions in
future publications.

The rest of the paper is organized as follows. In \refsec{methodology},
we derive the relation between the position-dependent power spectrum,
the squeezed-limit bispectrum, and the response of small-scale correlations
to large-scale overdensities. In \refsec{nbody}, we present measurements
of the position-dependent power spectrum from cosmological $N$-body
simulations. In \refsec{modeling}, we compare various theoretical
approaches to modeling the position-dependent power spectrum with
the simulations. We conclude in \refsec{conclusion}. In
\refapp{tr_bi_sq}, we derive the approximation of the squeezed-limit
tree-level matter bispectrum. 

\section{Position-dependent power spectrum, integrated bispectrum, and  linear response function}
\label{sec:methodology}
\subsection{Position-dependent power spectrum}
Consider a density fluctuation field, $\delta(\vr)$, in a {\it big} cubic
volume with the length of a side $L_B$. For simplicity and a
straightforward application to the $N$-body simulation box, we assume a
cubic volume and cubic subvolumes thereof. However, the method is also
applicable to realistic survey geometries without major changes. We
divide $L_B$ into $N_{\rm cut}$ pieces with the length of a side of each
{\it subvolume} given by $L=L_B/N_{\rm cut}$. In the subvolume centered
at $\vr_L$, we measure the local mean density perturbation relative to
the global mean density of the big volume, $\bar\delta(\vr_L)$, and the
position-dependent power spectrum, $P(\vk,\vr_L)$.  The local mean
overdensity within a subvolume centered at $\vr_L$ is given by
\begin{equation}
 \bar\delta(\vr_L) = \frac{1}{V_L}\int d^3r~\delta(\vr)W_L(\vr-\vr_L) ~,
\label{eq:deltabar}
\end{equation}
where $\delta(\vr)$ is the underlying overdensity relative to the global
mean density at a position $\vr$ and $V_L=L^3$ is the volume of the
subvolume. The window function is given by
\be
W_L(\vx)=\prod_{i=1}^3\:\theta(x_i), \quad
\theta(x_i) = \left\{
\begin{array}{cc}
1, & |x_i|\le L/2, \\
0,  & \mbox{otherwise}~.
\end{array}\right.
\label{eq:WL}
\ee
The Fourier transform is $W_L(\vq)=L^3\prod_{i=1}^3{\rm
sinc}(q_iL/2)$, where ${\rm sinc}(x)=\sin(x)/x$. 

We define the position-dependent power spectrum as
\be
P(\vk,\vr_L)\equiv \frac1{V_L}|\delta(\vk,\vr_L)|^2\,,
\ee 
where $\delta(\vk,\vr_L)\equiv \int_{V_L}d^3r~\delta(\vr)e^{-i\vr\cdot\vk}$
is the {\it local} Fourier transformation of the density fluctuation
field. The integral ranges over the subvolume centered at $\vr_L$. With
this quantity, the mean density perturbation in the subvolume centered
at $\vr_L$ is given by
$\bar\delta(\vr_L)=(1/V_L)\delta(\vk=0,\vr_L)$. One can use the window
function $W_L$ to extend the integration boundaries to infinity
\begin{equation}
 \delta(\vk,\vr_L)=\int d^3r~\delta(\vr)W_L(\vr-\vr_L)e^{-i\vr\cdot\vk}
 =\int\frac{d^3q}{(2\pi)^3}~\delta(\vk-\vq)W_L(\vq)e^{-i\vr_L\cdot\vq}~.
\label{eq:localdeltak}
\end{equation}
Therefore, the position-dependent power spectrum of the subvolume
centered at $\vr_L$ is 
\begin{equation}
 P(\vk,\vr_L)=\frac{1}{V_L}\int\frac{d^3q_1}{(2\pi)^3}\int\frac{d^3q_2}{(2\pi)^3}~
 \delta(\vk-\vq_1)\delta(-\vk-\vq_2)W_L(\vq_1)W_L(\vq_2)e^{-i\vr_L\cdot(\vq_1+\vq_2)} ~.
\label{eq:localpk}
\end{equation}

\subsection{Integrated bispectrum}
Correlating $P(\vk,\vr_L)$ with the local mean density perturbation of
the corresponding subvolume, we find
\begin{eqnarray}
 \langle P(\vk,\vr_L)\bar\delta(\vr_L)\rangle&=&
 \frac{1}{V_L^2}\int\frac{d^3q_1}{(2\pi)^3}\int\frac{d^3q_2}{(2\pi)^3}
 \int\frac{d^3q_3}{(2\pi)^3}~\langle\delta(\vk-\vq_1)
 \delta(-\vk-\vq_2)\delta(-\vq_3)\rangle \nonumber\\
 &&~~~~~~~~~~~~~~~~~~~~~~~~~~~
 \times W_L(\vq_1)W_L(\vq_2)W_L(\vq_3)e^{-i\vr_L\cdot(\vq_1+\vq_2+\vq_3)} ~,
\label{eq:corr_localpk_deltabar}
\end{eqnarray}
where $\langle\ \rangle$ denotes the ensemble average over many
universes. In the case of a simulation or an actual survey, the average
is taken instead over all the subvolumes in the simulation or the
survey volume. Through the definition of the bispectrum, 
$\langle\delta(\vq_1)\delta(\vq_2)\delta(\vq_3)\rangle=B(\vq_1,\vq_2,\vq_3)(2\pi)^3\delta_D(\vq_1+\vq_2+\vq_3)$
where $\delta_D$ is the Dirac delta function, \refeq{corr_localpk_deltabar} can be rewritten as
\begin{eqnarray}
\nonumber
 \langle P(\vk,\vr_L)\bar\delta(\vr_L)\rangle
 &=&\frac{1}{V_L^2}\int\frac{d^3q_1}{(2\pi)^3}\int\frac{d^3q_3}{(2\pi)^3}~
 B(\vk-\vq_1,-\vk+\vq_1+\vq_3,-\vq_3)\\
\nonumber
& &\hspace{3.3cm}\times W_L(\vq_1)W_L(-\vq_1-\vq_3)W_L(\vq_3) \\
&\equiv& iB(\vk)~.
\label{eq:iB}
\end{eqnarray}
As anticipated, the correlation of the position-dependent power spectrum
and the local mean density perturbation is given by an integral of the
bispectrum, and we will therefore refer to this quantity as the {\it
integrated bispectrum}, $iB({\bf k})$.

As expected from homogeneity, the integrated bispectrum is independent
of the location ($\vr_L$) of the 
subvolumes. Moreover, for an isotropic window function and bispectrum,
the result is also independent of the direction of $\vk$. The cubic
window function \refeq{WL} is of course not entirely spherically
symmetric,\footnote{We choose the cubic subvolumes merely for
simplicity. In general one can use any shapes. For 
example, one may prefer to divide the subvolumes into spheres, which naturally
lead to an isotropic integrated bispectrum $iB(k)$.} and there is a
residual dependence on $\hat{\vk}$ in \refeq{iB}. In the following, we
will focus on the angle average of \refeq{iB},
\begin{eqnarray}
 iB(k)&\equiv&\int\frac{d^2\Omega_{\hat\vk}}{4\pi}~iB(\vk)
 =\left\langle\left(\int\frac{d^2\Omega_{\hat\vk}}{4\pi}P(\vk,\vr_L)\right)\bar\delta(\vr_L)\right\rangle \nonumber\\
 &=&\frac{1}{V_L^2}\int\frac{d^2\Omega_{\hat\vk}}{4\pi}\int\frac{d^3q_1}{(2\pi)^3}
 \int\frac{d^3q_3}{(2\pi)^3}~B(\vk-\vq_1,-\vk+\vq_1+\vq_3,-\vq_3)\nonumber\\
& &\hspace{4.8cm}\times  W_L(\vq_1)W_L(-\vq_1-\vq_3)W_L(\vq_3) ~. 
\label{eq:iB_angular_average}
\end{eqnarray}
The integrated bispectrum contains integrals of three sinc
functions, ${\rm sinc}(x)$, which are damped oscillating functions and peak at
$|x|\lesssim \pi$. Most of the contribution to the integrated bispectrum thus
comes from values of $q_1$ and $q_3$ at approximately $1/L$. If the
wavenumber $\vk$ we are interested in is much larger than $1/L$ (e.g.,
$L=300~h^{-1}~{\rm Mpc}$ and $k\gtrsim0.3~h~{\rm Mpc}^{-1}$), then the dominant
contribution to the integrated bispectrum comes from the bispectrum in
squeezed configurations, i.e., $B(\vk-\vq_1,-\vk+\vq_1+\vq_3,-\vq_3) \to
B(\vk,-\vk,-\vq_3)$ with $q_1\ll k$ and $q_3\ll k$.

\subsection{Linear response function}
\label{sec:response}
Consider the following general separable bispectrum,
\be
B(\vk_1, \vk_2, \vk_3) = f(\vk_1,\vk_2) P(k_1) P(k_2) + 2 \:\rm perm~,
\label{eq:bis_gen}
\ee
where $f(\vk_1,\vk_2) = f(k_1, k_2, \hat{\vk}_1\cdot\hat{\vk}_2$) is a
dimensionless symmetric function of two $k$ vectors and the angle
between them. If $f$ is non-singular as one of the $k$ vectors goes
to zero, we can write, to lowest order in $q_1/k$ and $q_3/k$,
\ba
B(\vk-\vq_1,-\vk+\vq_1+\vq_3,-\vq_3) =\:& f(\vk-\vq_1,-\vq_3) P(|\vk-\vq_1|) P(q_3) \vs
& + f(-\vk+\vq_1+\vq_3,-\vq_3) P(|-\vk+\vq_1+\vq_3|) P(q_3) \vs
& + f(\vk-\vq_1,-\vk+\vq_1+\vq_3) P(|\vk-\vq_1|)P(|-\vk+\vq_1+\vq_3|) \vs
=\:& 2 f(\vk, 0) P(k) P(q_3) + f(\vk,-\vk) [P(k)]^2 + \cO\left(\frac{q_{1,3}}{k}\right)\,.
\ea
For matter, momentum conservation requires that $f(\vk,-\vk)=0$
\cite{peebles:1974}, as can explicitly be verified for the 
$F_2$ kernel of perturbation theory.
We then obtain
\be
 \int\frac{d^2\Omega_{\hat\vk}}{4\pi} B(\vk-\vq_1,-\vk+\vq_1+\vq_3,-\vq_3)
= \tilde f(k) P(k) P(q_3) + \cO\left(\frac{q_{1,3}}{k}\right)^2~,
\label{eq:bis_sq}
\ee
where $\tilde f(k) \equiv 2 f(0, k)$.  Note that the terms linear in $q_{1,3}$
cancel after angular average.  For a singular kernel, one has to
take into account the pole (see \refapp{tr_bi_sq}).  
Since the window function in real space
satisfies $W_L^2(\vr) = W_L(\vr)$, we have
$\int\frac{d^3q_1}{(2\pi)^3}~W_L(\vq_1)W_L(-\vq_1-\vq_3)=W_L(\vq_3)$.   
Performing the $\vq_1$ integral in \refeq{iB_angular_average} then yields
\ba
 iB(k) \stackrel{k L\to\infty}{=}\:& \frac{1}{V_L^2}  \int\frac{d^3q_3}{(2\pi)^3}
W_L^2(\vq_3) P(q_3) \tilde f(k) P(k)~. \nonumber\\
=\:& \sigma_L^2 \; \tilde f(k) P(k)~,
\label{eq:iB_sq}
\ea
where $\sigma_L^2$ is the variance of the density field on the
subvolume scale,
\be
\sigma_L^2 \equiv \frac{1}{V_L^2}  \int\frac{d^3q_3}{(2\pi)^3}
W_L^2(\vq_3)  P(q_3)~.
\label{eq:sigmaL}
\ee
\refEq{iB_sq} shows that the integrated bispectrum measures
how the small-scale power spectrum, $P(k)$, responds to a large-scale
density fluctuation with variance $\sigma_L^2$, with a response function
given by $\tilde{f}(k)$.

An intuitive way to arrive at the same expression is to
write the response of the small-scale power spectrum to a large-scale
density fluctuation as
\be
P(\vk,\vr_L) = \left.P(\vk)\right|_{\bar{\delta}=0} +
\left.\frac{dP(\vk)}{d\bar{\delta}}\right|_{\bar{\delta}=0}\bar{\delta}(\vr_L)+\dots~,  
\label{eq:Pkexp}
\ee
where we have neglected gradients and higher derivatives of $\bar\delta(\vr_L)$.  
We then obtain, to leading order,
\be
iB(k)=\sigma_L^2\left.\frac{d\ln
P(k)}{d\bar{\delta}}\right|_{\bar{\delta}=0}P(k). 
\label{eq:CZ}
\ee
Comparing this result with \refeq{iB_sq}, we find that $\tilde{f}(k)$
indeed corresponds to the {\it linear response} of the small-scale power to the
large-scale density fluctuation, $d\ln P(k)/\bar{\delta}$.  
In \refsec{bispectrum}, we use the full bispectrum of the form of
\refeq{bis_gen} and confirm the validity of the squeezed-limit result
given in \refeq{iB_sq}.  In \refsec{sep_uni}, we start from 
$d\ln P(k)/\bar{\delta}$ to compute $iB(k)$.
Inspired by \refeq{CZ}, we define another quantity, the {\it normalized integrated
bispectrum}, $iB(k)/[\sigma_L^2 P(k)]$. This quantity is equal to 
$\tilde{f}(k)$ and the linear response function in the limit of $kL\to \infty$.

\section{$N$-body simulations}
\label{sec:nbody}
We now present measurements of the position-dependent power spectrum from 160 collisionless
$N$-body simulations of a $2400~h^{-1}~{\rm Mpc}$ box with $768^3$ particles.
The same simulations are used in \cite{deputter/etal:2011}, and we refer to
section~3 of \cite{deputter/etal:2011} for more details. In short, the initial
conditions are set up using different realizations of Gaussian random fields
by second-order Lagrangian perturbation theory \cite{crocce/pueblas/scoccimarro:2006}
with the power spectrum given by \texttt{CAMB} \cite{lewis/bridle:2002}. We
adopt a flat $\Lambda$CDM cosmology, and the cosmological parameters are
$\Omega_m=0.27$, $\Omega_bh^2=0.023$, $h=0.7$, $n_s=0.95$, and
$\sigma_8=0.7913$.

To construct the density fluctuation field on grid points, we first distribute
all the particles in the $2400~h^{-1}~{\rm Mpc}$ box onto a $1000^3$ grid by the
cloud-in-cell (CIC) density assignment scheme. Then the density
fluctuation field at the grid point $\vr_g$ is
$\delta(\vr_g)=N(\vr_g)/\bar N-1$, where $N(\vr_g)$ 
is the fractional number of particles after the CIC assignment at $\vr_g$ and
$\bar N=768^3/1000^3$ is the mean number of particles in each grid cell.

We then divide the $2400~h^{-1}~{\rm Mpc}$ box in each dimension by
$N_{\rm cut}=4$, 8, and 20, so that there are 64, 512, and 8000
subvolumes with a side length of 600, 300, and $120~h^{-1}~{\rm Mpc}$,
respectively. The 
mean density perturbation in a subvolume centered at $\vr_L$ 
is $\bar\delta(\vr_L)=N_{\rm grid}^{-3}\sum_{\vr_g\in V_L}\delta(\vr_g)$, where
$(N_{\rm grid})^3=(1000/N_{\rm cut})^3$ is the number of grid points within
the subvolume. To compute the position-dependent power spectrum, we use
\texttt{FFTW}\footnote{Fast Fourier Transformation library:
\url{www.fftw.org}} to Fourier transform $\delta(\vr_g)$ in each
subvolume with the grid size $(N_{\rm grid})^3$. While the fundamental
frequency of the subvolume, $k_F=2\pi/L$, decreases with the subvolume
size $L$, the Nyquist frequency of the FFT grid, $k_{Ny}=k_FN_{\rm
grid}/2\approx 1.3 ~h^{-1}~{\rm Mpc}$, is the same in all cases. 

The position-dependent power spectrum is then computed as
\be
P(k,\vr_L)=\frac1{V_LN_{\rm mode}}
\sum_{k-\Delta k/2\le|\vk_i|\le k+\Delta
k/2}|\delta(\vk_i,\vr_L)|^2, 
\ee
where $N_{\rm mode}$ is the number of Fourier modes in the bin
$\left[k-\Delta k/2, k+\Delta k/2\right]$, and we set $\Delta
k\approx0.01~h~{\rm Mpc}^{-1}$ in all cases.  We choose this $\Delta k$
for all $N_{\rm cut}$ to sample well the baryon acoustic oscillations
(BAO) and thereby are able to show how the window function of the
different subvolumes damps the BAO (see below). We follow the
procedures in \cite{jing:2005} to correct for the the smoothing due to
the CIC density assignment and also for the aliasing effect in the power
spectrum. Note, however, that this correction is only important for
wavenumbers near the Nyquist frequency.

\begin{figure}[t]
\centering{
\includegraphics[width=1\textwidth]{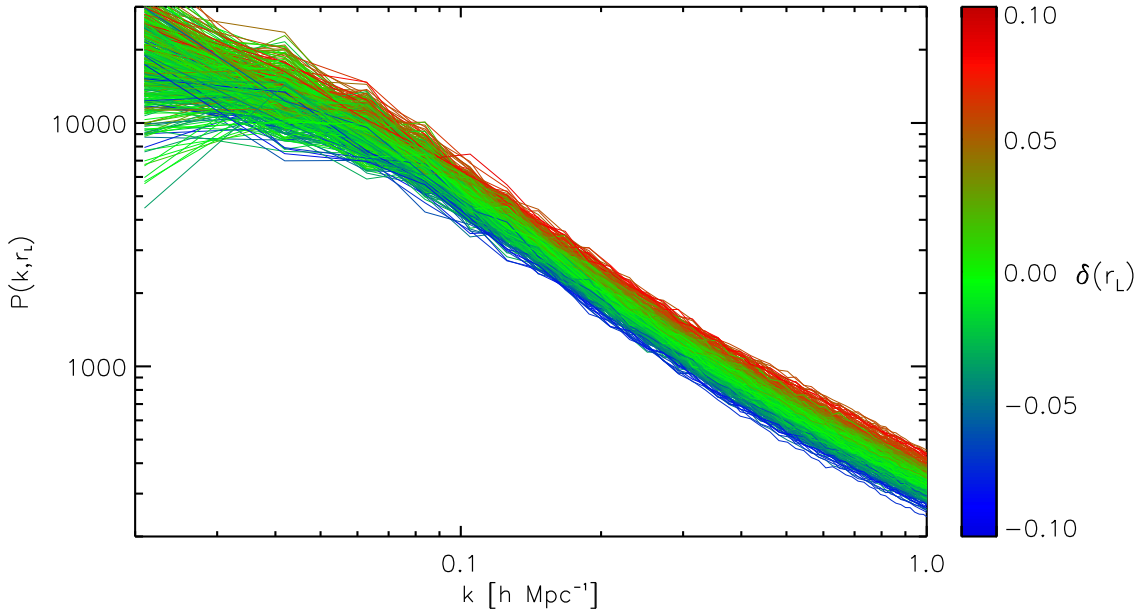}
}
\caption{Position-dependent power spectra measured from 512 subvolumes with
$L=300~h^{-1}~{\rm Mpc}$ in one realization. The color represents
 $\bar\delta(\vr_L)$ of each subvolume.
}
\label{fig:pk_dm_color}
\end{figure}

\refFig{pk_dm_color} shows the position-dependent power spectrum measured
from 512 subvolumes with $L=300~h^{-1}~{\rm Mpc}$ in one
realization. The color represents $\bar\delta(\vr_L)$ of each
subvolume. The positive correlation between the subvolume power spectra
and $\bar\delta(\vr_L)$ is obvious. The response $\tilde f(k) > 0$ is
clearly measurable at high significance in the simulations.

We measure the integrated bispectrum through
\be
iB(k)=\frac1{N_{\rm cut}^3}\sum_{i=1}^{N_{\rm cut}^3}P(k,\vr_{L,i})
\bar\delta(\vr_{L,i}),
\ee
where $P(k,\vr_{L,i})$ and $\bar\delta(\vr_{L,i})$ are measured in the
$i^{\rm th}$ subvolume. Further, motivated by \refeq{iB_sq}, we
normalize the integrated bispectrum by the mean power spectrum in the
subvolumes, $\langle P(k,\vr_L)\rangle=N_{\rm
cut}^{-3}\sum_{i=1}^{N_{\rm cut}^3}P(k,\vr_{L,i})$, and the variance of 
the mean density fluctuation in the subvolumes,
$\langle\sigma_L^2\rangle=N_{\rm cut}^{-3}\sum_{i=1}^{N_{\rm
cut}^3}\bar\delta^2(\vr_{L,i})$. This quantity,
$iB(k)/[\langle P(k,\vr_L)\rangle\langle\sigma_L^2\rangle]$, 
is the normalized integrated bispectrum we have defined at the end of
\refsec{response}, and is equal to the linear  response function, $d\ln
P(k)/d\bar\d$, given in \refeq{CZ} in the limit of $kL\to \infty$.

\begin{figure}[t]
\centering{
\includegraphics[width=1\textwidth]{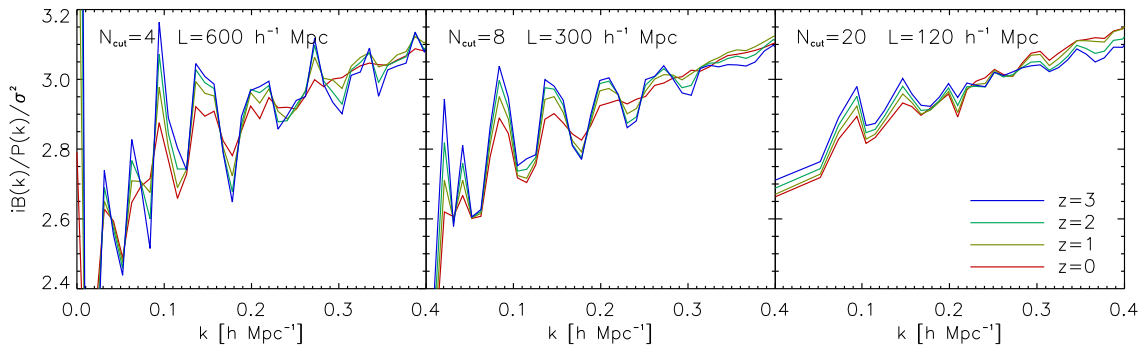}
}
\caption{Normalized integrated bispectrum, averaged over 160
 collisionless $N$-body simulations with Gaussian 
 initial conditions. From left to right 
 are $N_{\rm cut}=4$ ($L=600~h^{-1}~{\rm Mpc}$), 8 ($300~h^{-1}~{\rm Mpc}$),
 and 20 ($120~h^{-1}~{\rm Mpc}$); the blue, green, yellow, and red lines are
 $z=3,$ 2, 1, and 0, respectively. For clarity, we do not show the error
 bars. 
}
\label{fig:nbody_meas}
\end{figure}

\refFig{nbody_meas} shows the normalized integrated bispectrum,
averaged over 160 collisionless $N$-body simulations at different
redshifts. For clarity, no error bars are shown in this figure.
We have compared the results with a higher-resolution simulation with
$1536^3$ particles and starting at higher redshift ($z=49$ compared
to $z=19$ for our 160 simulations). For the scales and redshifts
shown in \reffig{nbody_meas}, the differences are less than 1\%.
However, we expect an up to 5\% uncertainty in the integrated bispectrum
at $z=3$ (less at lower $z$) due to transients which affect the
bispectrum more strongly than the power spectrum \cite{mccullagh/jeong},
as well as other systematics such as mass resolution.

Since the initial conditions are Gaussian, the bispectrum is generated
entirely by non-linear gravitational evolution. We thus measure
the effect of a long-wavelength density perturbation on the evolution of
small-scale structures.  
The wiggles visible in each panel of \reffig{nbody_meas} are due to the
BAOs. The BAOs in the right panel are strongly damped because the box
size ($120~h^{-1}~{\rm Mpc}$) approaches the BAO scale, and the
window function smears the BAO feature \cite{chiang/etal:2013}.  
Further, BAO amplitudes are larger at higher redshifts as they are
less damped by non-linear evolution \cite{eisenstein/seo/white:2007}.  
The broad-band shape of the normalized integrated bispectrum evolves on
small scales due to non-linear evolution, leading to an
effective steepening of its slope. We now turn to the
theoretical modeling of the results shown in \reffig{nbody_meas}.

\section{Theoretical modeling}
\label{sec:modeling}
We use two different approaches to model the integrated bispectrum. 
In the first approach, we model the bispectrum and compute the integral
to obtain the integrated bispectrum. In the second approach, we model
the response of the small-scale power spectrum to a long wavelength
perturbation directly using the ``separate universe'' picture.  
For clarity, we will show the comparison between model prediction
and simulations only for the $L=300~h^{-1}~{\rm Mpc}$ subvolumes 
($N_{\rm cut}=8$).  The agreement with simulations is independent of
subvolume size as long as the subvolume size is large enough for
$\bar{\delta}$ to be in the linear regime, and the window function is
taken into account.

\subsection{Bispectrum modeling}
\label{sec:bispectrum}
We compute the integrated bispectrum by using a model for the bispectrum
in \refeq{iB_angular_average} and perform the eight-dimensional
integral. Because of the high dimensionality, we use the Monte Carlo
integration routine in GNU Scientific Library to evaluate the angular-averaged integrated bispectrum.
In the following, we consider two different models for the matter bispectrum. 

\subsubsection{Standard perturbation theory}
\label{sec:f2_spt}
The standard perturbation theory (SPT) \cite{bernardeau/etal:2001} gives
the tree-level matter bispectrum as 
\be
B_{\rm SPT}(\vk_1,\vk_2,\vk_3)=2[P_l(k_1)P_l(k_2)F_2(\vk_1,\vk_2)
+2 \:\rm perm], 
\label{eq:Bspt}
\ee
where $P_l(k)$ is the linear matter power spectrum, and
\begin{equation}
 F_2(\vk_1,\vk_2)=\frac{5}{7}+\frac{1}{2}\frac{\vk_1\cdot\vk_2}{k_1k_2}
 \left(\frac{k_1}{k_2}+\frac{k_2}{k_1}\right)+\frac{2}{7}
 \left(\frac{\vk_1\cdot\vk_2}{k_1k_2}\right)^2 ~.
\end{equation}
In order to normalize the integrated bispectrum, we need an expression
for the mean subvolume power spectrum $\langle P_{L} \rangle$.  For this
we use the linear power spectrum convolved with the window function,
\begin{equation}
 \langle P_{L,l}(k)\rangle=\frac{1}{V_L}\int\frac{d^3q}{(2\pi)^3}~
 P_l(|\vk-\vq|)|W_L(\vq)|^2 ~,
\label{eq:pk_modeling}
\end{equation}
while the variance of the mean density fluctuation in the subvolumes is
given by \refeq{sigmaL}. Both quantities are calculated through Monte
Carlo integration.

\begin{figure}[t]
\centering{
\includegraphics[width=1\textwidth]{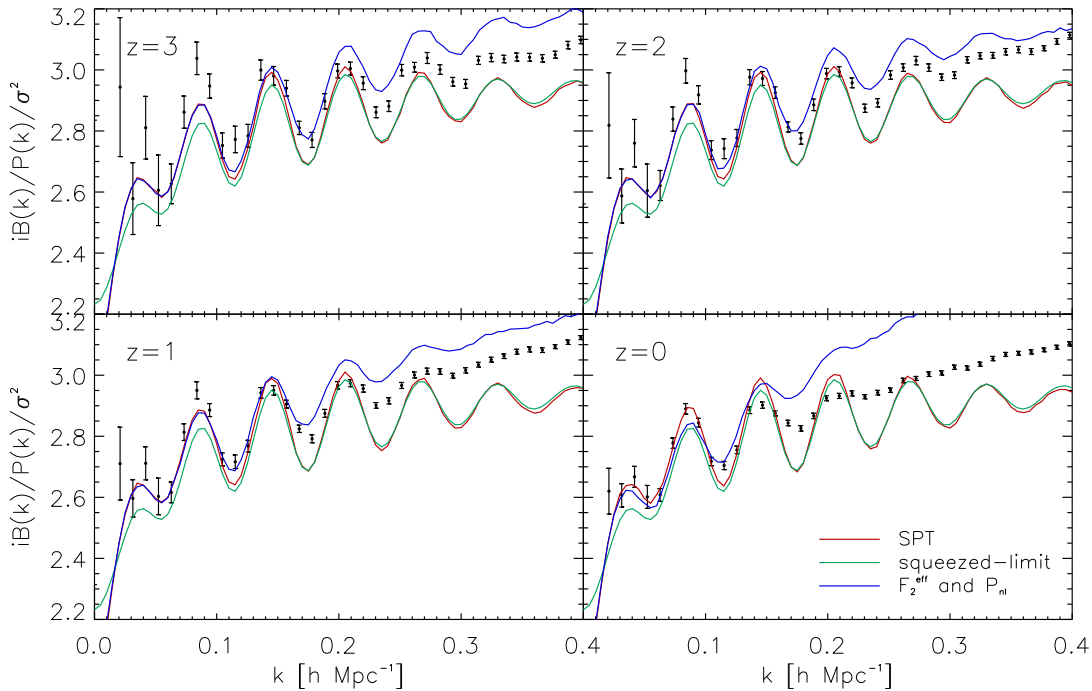}
}
\caption{The SPT and the $F_2^{\rm eff}(\vk_1,\vk_2)$ predictions for the
 normalized integrated  bispectrum at different redshifts. The red and
 blue lines  are computed by the direct integration of the eight-dimensional
 integral (\refeq{iB_angular_average}) with the standard
 $F_2(\vk_1,\vk_2)$ kernel and the linear power spectrum, and $F_2^{\rm
 eff}(\vk_1,  \vk_2)$ and the non-linear power spectrum, respectively.
 The green lines show the squeezed-limit approximation
 (\refeq{int_bi_approx}) to the SPT results. The $N$-body simulation
 results are shown by the black data points with the error bars showing
 the standard deviation on the mean measured from 160 simulations.
}
\label{fig:spt_f2eff_modeling}
\end{figure}

We compare the normalized integrated bispectrum measured from the
simulations with the SPT prediction in \reffig{spt_f2eff_modeling} (red
lines). The SPT prediction is independent of redshift. This is because 
the linear power spectra at
various redshifts are only different by the wavenumber-independent
linear growth factor, $D^2(z)$. Therefore, the linear growth factor
cancels out in the normalized integrated bispectrum. The SPT predictions
agree with the simulations relatively well at $z\ge1$ and
$k\lesssim0.2~h~{\rm Mpc}^{-1}$, whereas they fail at lower redshifts as
well as on smaller scales, where non-linearities become too strong to be
described by SPT. Especially, the BAO amplitudes at  $k\gtrsim0.2~h~{\rm
Mpc}^{-1}$ are affected: while the SPT predictions are
redshift-independent, the simulations show smaller BAO amplitudes at lower
redshifts. 

The eight-dimensional integral in \refeq{iB_angular_average} simplifies
greatly if we focus on the squeezed-limit bispectrum. In
\refapp{tr_bi_sq}, we show  
\begin{eqnarray}
\nonumber
& & \int\frac{d^2\Omega_{\hat\vk}}{4\pi}~B_{\rm
SPT}(\vk-\vq_1,-\vk+\vq_1+\vq_3,-\vq_3) \\
&=&
 \left[\frac{68}{21}-\frac{1}{3}\frac{d\ln k^3 P_l(k)}{d\ln k}\right]P_l(k)P_l(q_3) 
 + \cO\left[\left(\frac{q_{1,3}}{k}\right)^2\right] ~,
\label{eq:bi_approx}
\end{eqnarray}
for  $k\gg q_1,q_3$. This result has the same form as given in
\refeq{bis_sq}. We can then apply \refeq{iB_sq} and perform
all the integrals analytically in the limit of $kL\to \infty$ to obtain
\begin{eqnarray}
 iB_{\rm SPT}(k)&=&\frac{1}{V_L^2}\int\frac{d^2\Omega_{\hat\vk}}{4\pi}
 \int\frac{d^3q_1}{(2\pi)^3}\int\frac{d^3q_3}{(2\pi)^3}~
 B_{\rm SPT}(\vk-\vq_1,-\vk+\vq_1+\vq_3,-\vq_3) \nonumber\\
& &\hspace{4.8cm}\times W_L(\vq_1)W_L(-\vq_1-\vq_3)W_L(\vq_3)\nonumber\\
 &\stackrel{k L \to \infty}{=}&
 \left[\frac{68}{21}-\frac{1}{3}\frac{d\ln k^3 P_l(k)}{d\ln k}\right]P_l(k)\sigma_L^2 ~.
\label{eq:int_bi_approx}
\end{eqnarray}
Comparing this result with \refeq{CZ}, we find that
the linear response of the power spectrum in SPT is given by
\be
\left.\frac{d\ln P_l(k)}{d\bar{\delta}}\right|_{\rm SPT}
=\frac{68}{21}-\frac{1}{3}\frac{d\ln k^3 P_l(k)}{d\ln k}.
\label{eq:sptresponse}
\ee

The green lines in \reffig{spt_f2eff_modeling} show the squeezed-limit
approximation given in
\refeq{int_bi_approx}. While they are different from 
the full integration (red lines) at $k\lesssim0.2~h~{\rm Mpc}^{-1}$, for
which the squeezed-limit approximation fails and the direct integration
is required, they agree well---with the fractional difference
being less than 1.5\% (1\% for $L=600\, h^{-1}$ Mpc)---at
$k\gtrsim0.2~h~{\rm Mpc}^{-1}$, corresponding to a value of $1/(k L)
\lesssim 0.02$. Thus, the squeezed-limit is reached already with good precision
for $kL\gtrsim 50$. 

\refEq{int_bi_approx} does not contain
any window function effect apart from that in the variance $\sigma_L^2$. 
While this is a good approximation for the slowly-varying
part of the integrated bispectrum, it does not capture the smearing of the
BAO features due to the window function. We incorporate this effect by
replacing $d\ln P_l(k)/d\ln k$ with appropriately convolved forms, ${\rm
conv}[d P_l(k)/d\ln k]\:/\:{\rm conv}[P_{l}(k)]$, in \refeq{int_bi_approx}. This form 
is motivated by the separate universe approach discussed in
\refsec{sep_uni}, and provides an accurate result as shown in
\reffig{spt_f2eff_modeling}.

\subsubsection{Bispectrum fitting formula}
\label{sec:f2_eff}
The SPT predictions fail on smaller scales as well as at lower redshifts where
non-linearity becomes too strong to be described by SPT. 
An empirical fitting formula for non-linear evolution of the matter
bispectrum was proposed in \cite{scoccimarro/couchman:2000} and further
improved in \cite{gilmarin/etal:2011}. 
In short, the form is the same as the tree-level matter bispectrum, but
$F_2(\vk_1,\vk_2)$ is replaced by an effective kernel, $F_2^{\rm
eff}(\vk_1,\vk_2)$, which contains nine fitting parameters, $\{a_1,
\cdots  a_9\}$, to account for non-linearity (see eqs. 2.6 and 2.12 in
\cite{gilmarin/etal:2011} for details). Therefore, we use $F_2^{\rm
eff}(\vk_1,\vk_2)$ and compute the integrated bispectrum by performing
the eight-dimensional integral numerically with Monte Carlo
integration. We use the same values of 
the best-fit parameters provided in table 2 of
\cite{gilmarin/etal:2011}, which were calibrated by
fitting to simulation results between $z=0$ and $z=1.5$. In contrast to
the SPT formalism that uses the linear power spectrum in \refeq{Bspt},
the fitting formula uses the non-linear power spectrum, for which we use
the mean power spectrum measured from the 160 simulation boxes. For the
normalization of the integrated bispectrum, we convolve the non-linear
power spectrum with the subvolume window function as in \refeq{pk_modeling}.
Note that the $F_2^{\rm eff}$ fitting formula is not specifically
designed for the squeezed configuration, but instead was calibrated to
a wide range of triangle configurations of the matter bispectrum. 

The blue lines in \reffig{spt_f2eff_modeling} show the normalized
integrated bispectrum computed with $F_2^{\rm eff}$, which clearly depends
on redshift. At $z\gtrsim1$, the $F_2^{\rm eff}$ modeling and the simulations
are in good agreement at $k\lesssim0.2~h~{\rm Mpc}^{-1}$. At $k>0.2~h~{\rm Mpc}^{-1}$,
although the $F_2^{\rm eff}$ modeling predicts larger broad-band power
of the normalized integrated bispectrum, the BAO amplitudes still agree
well with the simulations. This is most obvious for the two BAO
peaks at $0.25~h~{\rm Mpc}^{-1}\le k\le0.35~h~{\rm Mpc}^{-1}$. On the
other hand, at $z=0$, the $F_2^{\rm eff}$ modeling predicts much larger
normalized integrated bispectrum on small scales than measured in the
simulations,  so that  the fitting formula does not perform much
better than tree-level perturbation theory at $z=0$.

\subsection{Separate universe approach}
\label{sec:sep_uni}
In the second approach, we compute the effects of a long-wavelength
density fluctuation on the small-scale power spectrum by treating each
over- and under dense region as a separate universe with a different
background density. This approach thus neglects the finite size of the
subvolumes and is valid for wavenumbers which satisfy $k L \gg 1$ 
(specifically, $k L \gtrsim 50$ for percent-level accuracy).

The power spectrum in a separate universe with an infinite-wavelength
density perturbation, $\bar\d$, with respect to the global flat
$\Lambda$CDM cosmology can be expanded as in \refeq{Pkexp}. Through
\refeq{CZ}, the normalized integrated bispectrum is equal to the linear response
of the non-linear matter power spectrum at wavenumber $k$ to $\bar\d$:
\be
\frac{iB(k)}{P(k) \sigma_L^2} = \frac{d \ln P(k)}{d \bar\d}\,.
\label{eq:iB_resp}
\ee
This is not exactly true if the subvolumes for which $iB(k)$ is
measured are not spherical. For example, since the cubic window function
is anisotropic, the integrated bispectrum might pick up contributions
from the tidal field. However, we have verified that the anisotropy of
the cubic window function has a negligible effect, by computing the
dipole and quadrupole of the integrated bispectrum through
\refeq{iB_angular_average}. The ratios to the monopole are less than $10^{-5}$ 
on the scales of interest.

A universe with an infinite-wavelength density perturbation with respect
to a flat fiducial cosmology is equivalent to a universe with non-zero
curvature (e.g., \cite{baldauf/etal:2011}). This alters the scale
factor-time relation, Hubble rate, and linear growth, and thus affects
the power spectrum. Recent papers 
\cite{creminelli/etal:2013,valageas:2013,kehagias/perrier/riotto:2013,li/hu/takada:2014}
have studied this topic. We briefly summarize the result here. We write
the fractional mass density perturbation with respect to the
fiducial flat universe as 
\be
\bar\d(t) = \frac{\tilde{\rhob}(t)}{\rhob(t)} - 1 = \frac{D(t)}{D(t_0)} \bar\d_0\,,
\ee
where $\rhob(t)$ is the background matter density in the fiducial
flat cosmology, $D(t)$ is the linear growth factor in the
same cosmology, $\tilde{\rhob}$ is the background matter density
in a slightly curved universe,  $t_0$ is a reference time, and $\bar\d_0$
is the density perturbation at 
$t_0$. In the following, $t_0$ will stand for the present epoch and we
choose $a(t_0)=1$. Quantities in the modified (curved)
cosmology are denoted with a tilde, such as $\tilde a(t)$ for the
modified scale factor-time relation. Note that the time coordinates are
the same in both cosmologies in the sense that they are the proper time
for comoving observers in the absence of perturbations. For a fiducial
flat $\Lambda$CDM cosmology, the modified curved cosmology to linear
order in $\bar\d$ is described by modified cosmological parameters (also see
\cite{sirko:2005,li/hu/takada:2014})
\ba
\tilde H_0 =\:& H_0\left( 1 + \d_H \right)
\vs
\tOm =\:& \Om \left(1 - 2\d_H \right)
\vs
\tOK =\:& -\left(\Om  + \frac23 f_0 \right)\bar\d_0
\vs
\tilde{\Omega}_{\Lambda} =\:& \Omega_\Lambda \left(1 - 2\d_H\right)\,,
\label{eq:mod_param}
\ea
where
\be
\d_H \equiv \left(-\frac12 \Om - \frac13 f_0 \right) \bar\d_0,
\ee
and $f_0 = f(t_0)$ is the logarithmic growth rate evaluated at $t_0$.  The scale 
factor-time relation in the modified cosmology is given by
\be
\tilde a(t) = a(t) \left[1 - \frac13 \bar\d(t) \right]\,.
\ee
Hence, observables calculated with respect to comoving coordinates in the 
modified cosmology $\tilde a(t)$ have to be transformed
according to a coordinate rescaling of
\be
\vx \to \vx' = \vx\left[1 - \frac13 \bar\d(t) \right]\,.
\label{eq:xresc}
\ee
For the power spectrum, this corresponds to (see appendix~A of
\cite{conformalfermi})
\be
P(k,t) \to P(k,t) \left[1 - \frac13 \frac{d\ln k^3 P(k,t)}{d\ln k}
\bar\d(t) \right]\,.
\label{eq:fermi}
\ee

Let us denote the power spectrum in the modified cosmology described by
\refeq{mod_param} as $\tilde P(\tilde k, t)$. This power
spectrum refers to the modified mean density, which is
given by the fiducial mean density multiplied 
by $1 + \bar\d(t)$. We then have for the power
spectrum with respect to the fiducial mean density
\be
 P(\tilde{k},t) =\left[1 + 2\bar\d(t)\right]
 \tilde{P}(\tilde{k},t)\,. 
\ee
Converting $\tilde k$ to $k$ with \refeq{fermi}
and using  the scale factor instead of time, the  power spectrum in the
presence of $\bar\d$ is given by
\be
P(k,a | \bar\d) = \tilde P\left(k, a\left[1-\frac13 \bar\d(a)\right]\right) 
\left[1 + \left(2 - \frac13\frac{d\ln k^3 P(k,a)}{d\ln k}\right)  \bar\d(a)\right]~.
\label{eq:Pkd0}
\ee
Note that this expression is only valid to linear order in $\bar\d$.  

Both $P(k)$ and $\bar\d$ are measured in a finite volume,
described by the window function $W_L$. In order to take this into
account, \refeq{Pkd0} is convolved by the window function.  Note that we
take the convolution \emph{after} applying the derivative $d\ln
k^3P(k)/d\ln k$, rather than taking the derivative of the convolved
power spectrum. This is because the window function is fixed in terms of
observed coordinates  (in the fiducial cosmology), i.e., it is not
subject to the rescaling of \refeq{xresc}. Taking the slope of the
convolved power spectrum would correspond to a window function defined
in the ``local'' curved cosmology. 

\subsubsection{Linear power spectrum}
\label{sec:linearresponse}
For the linear power spectrum, $P_l$, we have
\be
\tilde P_l\left(k, a\left[1-\frac13 \bar\d(a)\right]\right)
= \left(\frac{\tilde D\left(a\left[1-\frac13  \bar\d(a)\right]\right)}{D(a)}\right)^2 P_l(k, a)\,.
\ee
The linear growth factor is changed following  (see appendix D
in \cite{baldauf/etal:2011})
\be
\tilde D\left(a\left[1-\frac13 \bar\d(a)\right]\right)
= D(a) \left[1 + \frac{13}{21} \bar\d(a) \right]\,,
\label{eq:Dtilde}
\ee
where $D(a)$ is the growth factor in the fiducial cosmology. The
prefactor $13/21$ is only strictly valid for an Einstein-de Sitter
cosmology; however, the cosmology dependence is very mild. The
fractional difference of $d\ln D(a)/d\bar\d$ between $\Lambda$CDM
cosmology and Einstein-de Sitter universe at $z=0$ is at the 0.1\%
level. 

Putting everything together, \refeq{Pkd0} yields for the linear response
function of the linear power spectrum 
\begin{equation}
\frac{d \ln P_l(k, a)}{d \bar\d(a)} = 
\frac{68}{21} - \frac13 \frac{d\ln k^3 P_l(k,a)}{d\ln k}\,.
\label{eq:Pkd0L}
\end{equation}
This result (which again is only exact for Einstein-de Sitter) matches
the expression derived from the $F_2$ kernel given in \refeq{sptresponse}.

\subsubsection{SPT 1-loop power spectrum}
\label{sec:1loopresponse}

Expanding matter density fluctuations to third order, one obtains the
so-called ``SPT 1-loop power spectrum'' given by
$P(k,a)=P_l(k,a)+P_{22}(k,a)+2P_{13}(k,a)$, where \cite{bernardeau/etal:2001}
\begin{eqnarray}
\label{eq:P22}
P_{22}(k,a) &=& 
2 \int \frac{d^3 q}{(2\pi)^3} P_{l}(q,a) P_{l}(|\mathbf{k}-\mathbf{q}|,a)
            \left[F_2(\mathbf{q},\mathbf{k}-\mathbf{q})\right]^2,\\
 \nonumber
   2P_{13}(k,a) &=&
      \frac{2\pi k^2}{252} P_{l}(k,a) \int_{0}^{\infty} \frac{dq}{(2\pi)^3}
      P_{l}(q,a)\\
\nonumber
& & \times \Biggl[100\frac{q^2}{k^2} -158 + 12\frac{k^2}{q^2} 
   -42 \frac{q^4}{k^4}
+\frac{3}{k^5q^3}(q^2-k^2)^3(2k^2+7q^2)\ln\left( \frac{k+q}{|k-q|} \right)
      \Biggl].
\label{eq:P13}
\end{eqnarray}
Both $P_{22}$ and $P_{13}$ are proportional to $D^4(a)$. Modifying the
growth factor as described in \refsec{linearresponse}, we obtain the linear
response function of the SPT 1-loop power spectrum as
\be
\frac{d \ln P(k, a)}{d \bar\d(a)} = 
 \frac{68}{21} - \frac13 \frac{d\ln k^3 P(k,a)}{d\ln
 k}+\frac{26}{21}\frac{P_{22}(k,a)+2P_{13}(k,a)}{P(k,a)}\,.
\label{eq:Pkd01loop}
\ee
Note that this can easily be generalized to $n$ loops in perturbation
theory by using that $d\ln P_{(n-{\rm loop})}(k,a)/d\ln D(a) = 2n+2$.  
We include the window function effect by computing  ${\rm
conv}[dP(k)/d\bar{\delta}]/{\rm conv}[P(k)]$.

\begin{figure}[t]
\centering{
\includegraphics[width=1\textwidth]{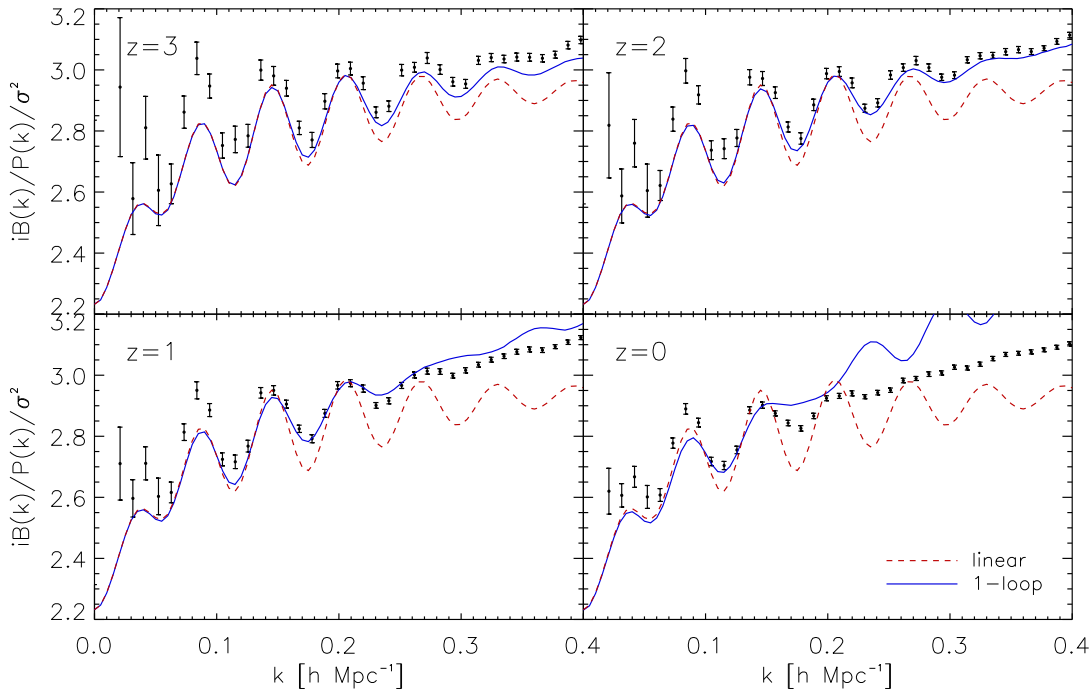}
}
\caption{Normalized integrated bispectrum from the $N$-body simulations
 (points with error bars) and the linear response functions, $d\ln P(k,
 a)/d\bar\d(a)$, computed from the separate universe approach
 combined with perturbation theory. The dashed lines
 show the linear theory results (\refeq{Pkd0L}), while the solid lines
 show the SPT 1-loop results (\refeq{Pkd01loop}). The agreement between
 the 1-loop predictions and the simulation results is very good at $z\ge
 1$. Note that the difference between the normalized integrated
 bispectrum and the linear response function at $k\lesssim 0.2~h~{\rm
 Mpc}^{-1}$ is due to the squeezed limit not being reached yet (see the
 text below \refeq{sptresponse}).
}
\label{fig:sep_uni_pert}
\end{figure}

\refFig{sep_uni_pert} compares the linear theory and the SPT
1-loop predictions with the $N$-body simulation results.  
The SPT 1-loop prediction captures the damping of BAOs due to non-linear
evolution, and agrees well with the simulation results at $z=1$, 2, and 3.  
This is expected from the excellent performance of the 1-loop matter power
spectrum at high redshifts as demonstrated by \cite{jeong/komatsu:2006}.   
The agreement degrades rapidly at $z=0$, also as expected.  
Note that comparing $z=2$ and 3, the 1-loop prediction seems to agree
better with the measurements at $z=2$.  However, as mentioned in
\refsec{nbody}, transients and other systematics might have an impact of up
to 5\% on the measurements at $z=3$, which is larger than the difference
shown in the top left panel of \reffig{sep_uni_pert}.

\subsubsection{\texttt{halofit} and Coyote emulator}
We now apply the separate universe approach to simulation-calibrated
fitting formulae for the non-linear matter power spectrum, specifically
the \texttt{halofit} prescription
\cite{smith/etal:2002} and the Coyote emulator \cite{heitmann/etal:2013}.  
These prescriptions yield $\tilde P(k, a)$ for a given set of cosmological
parameters, so that \refeq{Pkd0} can be immediately applied.  However,
the Coyote emulator does not provide predictions for curved cosmologies, and
we hence adopt a simpler approach here.  

In case of the linear power spectrum, the effect of the modified cosmology
enters only through the modified growth factor given in \refeq{Dtilde}.
Correspondingly, we can approximate the effect on the non-linear power
spectrum by a change in the value of the power spectrum normalization
$\sigma_8$ at redshift zero, 
\be
\sigma_8 \to \left[1 + \frac{13}{21} \bar\d_0 \right] \sigma_8~,
\label{eq:sig8_resc}
\ee
where we have used the Einstein-de Sitter prediction. Therefore, the
non-linear power spectrum response becomes 
\begin{equation}
 \frac{d\ln P_{nl}(k,a)}{d\bar\d(a)}= \frac{13}{21}\frac{d\ln P_{nl}(k,a)}{d\ln \sigma_8}
 +2 - \frac{1}{3}\frac{d\ln k^3P_{nl}(k,a)}{d\ln k} ~.
\label{eq:Pkd0NL}
\end{equation}

\begin{figure}[t]
\centering{
\includegraphics[width=1\textwidth]{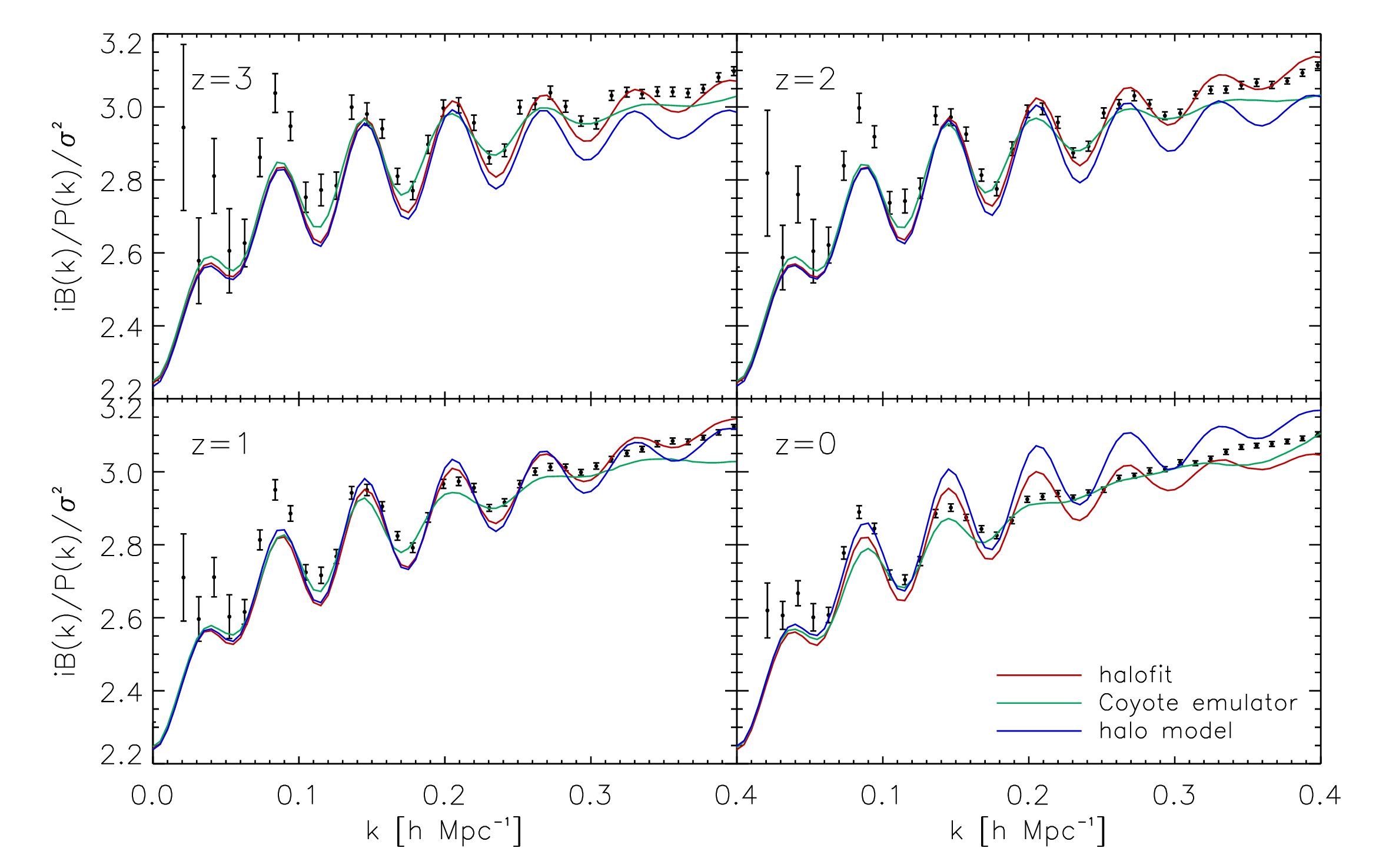}
}
\caption{Same as \reffig{sep_uni_pert}, but for the
 linear response functions computed from \texttt{halofit} (red), the Coyote
 emulator (green), and the halo model (blue).
}
\label{fig:sep_uni_nonpert}
\end{figure}

The results of applying \refeq{Pkd0NL} to \texttt{halofit} (red) and the
Coyote emulator (green) are shown in \reffig{sep_uni_nonpert}.  
In terms of broad-band power, the \texttt{halofit} prediction provides
a good match. However, the predicted BAO amplitude are larger than the 
measurement, especially at low redshift at $k\gtrsim0.3~h~{\rm Mpc}^{-1}$.
Also, while the BAO phases of \texttt{halofit} follow the SPT prediction,
there are some differences with respect to the measurement of the
$N$-body simulations due to the non-linear evolution.  
The Coyote emulator performs to better than $\sim 2$\% over
the entire range of scales and redshifts. It slightly underpredicts the 
small-scale power at $k > 0.3~h~{\rm Mpc}^{-1}$ for $z \geq 1$.
For redshifts $z \geq 2$ and on the scales considered,
the 1-loop predictions are of comparable accuracy to the Coyote emulator, 
while the latter provides a better fit at lower redshifts.  
Finally, note also our previous caveat regarding transients at the end of 
\refsec{1loopresponse}.

\subsubsection{Halo model}

In the halo model (see \cite{cooray/sheth} for a review), all matter is
assumed to be contained within halos with a certain distribution of mass
given by the mass function, and a certain density profile. Along with
the clustering properties of the halos, these quantities then determine
the statistics of the matter density field on all scales including the
non-linear regime. $N$-point functions can be conveniently decomposed
into 1- through $N$-halo pieces. In the following, we will follow the
most common halo model approach and assume a linear local bias of the
halos.

Adopting the notation of \cite{takada/hu:2013}, the halo model power
spectrum, $P_{\rm HM}(k)$, is given by
\ba
P_{\rm HM}(k) =\:& P^{\rm 2h}(k) + P^{\rm 1h}(k) \label{eq:PkHM}\\
P^{\rm 2h}(k) =\: & \left[I^1_1(k)\right]^2 P_l(k) \vs
P^{\rm 1h}(k) =\: & I^0_2(k,k)\,,\nonumber
\ea
where
\be
I^n_m(k_1,\cdots k_m) \equiv \int d\ln M\:n(\ln M) \left(\frac{M}{\rhob}\right)^m \, b_n(M)\, u(M|k_1) \cdots u(M|k_m)\,,
\label{eq:Inmdef}
\ee
and $n(\ln M)$ is the mass function (comoving number density per interval
in log mass), $M$ is the halo mass, $b_n(M)$ is the $n$-th order local bias 
parameter, and $u(M|k)$ is the dimensionless Fourier transform of the halo 
density profile, for which we use the NFW profile \cite{NFW}. We normalize
$u$ so that $u(M|k\to 0)=1$.  The notation given in \refeq{Inmdef}
assumes $b_0\equiv1$. $u(M|k)$ depends on $M$ through the scale radius
$r_s$, which in turn is given through the mass-concentration
relation. All functions of $M$ in \refeq{Inmdef} are also functions of
$z$ although we have not shown this for clarity. In the following,
we adopt the Sheth-Tormen mass function \cite{sheth/tormen} with
the corresponding peak-background split bias, and the mass-concentration
relation of \cite{bullock/etal}. The exact choice of the latter has negligible
impact on the mildly non-linear scales considered in this paper.

We now derive how the power spectrum given in \refeq{PkHM} responds to an
infinitely long-wavelength density perturbation $\bar\d$, as was done
for the \texttt{halofit} 
and Coyote emulator approaches. For this, we consider the 1-halo and 2-halo
terms separately. The key physical assumption we make is that halo profiles
in \emph{physical} coordinates are unchanged by the long-wavelength density
perturbation. That is, halos at a given mass $M$ in the presence of $\bar\d$
have the same scale radius $r_s$ and scale density $\rho(r_s)$ as in the
fiducial cosmology. This assumption, which is related to the stable clustering
hypothesis, can be tested independently with simulations, but this goes beyond
the scope of this paper. Given this assumption, the density perturbation
$\bar\d$ then mainly affects the linear power spectrum, which determines
the halo-halo clustering (2-halo term), and the abundance of halos at a
given mass. 

We begin with the 2-halo term. The response of the linear power spectrum
is given by \refeq{Pkd0L}. The expression for the 2-halo term in
\refeq{PkHM} is simply the convolution (in real space) of the halo
correlation function in the linear bias model with the halo density
profiles. By assumption, the density profiles do 
not change, hence $I^1_1$ only changes through the bias $b_1(M)$ and the mass
function $n(\ln M)$. The bias $b_N(M)$ quantifies the $N$-th order response
of the mass function $n(\ln M)$ to an infinite-wavelength density perturbation  \cite{mo/white:1995,PBSpaper}:
\be
b_N(M) = \frac{1}{n(\ln M)} \frac{\partial^N n(\ln M)}{\partial \bar\d^N}\Big|_0\,,
\label{eq:bNdef}
\ee
We then have
\ba
\frac{\partial n(\ln M)}{\partial\bar\d}\Big|_0 =\:& b_1(M) n(\ln M)\,, \vs
\frac{\partial b_1(M)}{\partial\bar\d}\Big|_0 =\:& - [b_1(M)]^2 + b_2(M)\,.
\ea
Thus,
\ba
\frac{\partial}{\partial\bar\d} I^1_1(k) =\:& \int d\ln M \:n(\ln M) \left(\frac{M}{\rhob}\right)
\left([b_1(M)]^2 - [b_1(M)]^2 + b_2(M) \right) u(M|k) \vs
=\:& \int d\ln M \:n(\ln M) \left(\frac{M}{\rhob}\right) b_2(M) u(M|k) \vs
=\:& I^2_1(k)\,.
\label{eq:dI11}
\ea
In the large-scale limit, $k\to 0$, this vanishes by way of the halo
model consistency relation
\be
\int d\ln M\: n(\ln M)\left(\frac{M}{\rhob}\right) b_N(M) = 
\left\{
\begin{array}{ll}
1, & N = 1\,, \\
0, & N \geq 1\,.
\end{array}\right.
\ee
For finite $k$ however, \refeq{dI11} does not vanish. Thus, the linear
response function of the two-halo term becomes 
\be
\frac{d P^{\rm 2h}(k)}{d\bar\d}\Big|_0 = \left[ \frac{68}{21}
- \frac13 \frac{d\ln k^3 P_l(k)}{d\ln k} \right] P^{\rm 2h}(k)
+ 2 I^2_1(k) I^1_1(k) P_l(k)\,.
\label{eq:Pk2hd0}
\ee
Note that we recover the tree-level result given in \refeq{Pkd0L} in the
large-scale limit. Strictly speaking, this expression is not consistent,
since the term $I^2_1$ implies a non-zero $b_2$ while in \refeq{PkHM} we
have assumed a pure linear bias. Of course, if we allowed for $b_2$ in
\refeq{PkHM}, we would obtain a contribution from $b_3$ in
\refeq{Pk2hd0}, and so on. This reflects the fact that the halo model
itself cannot be made entirely self-consistent. Note that in
\refeq{Pk2hd0} the slope is taken from the \emph{linear}, not 2-halo
power spectrum. This is a consequence of our assumption that halo
profiles do not change due to $\bar\d$; in other words, having $d\ln k^3
P^{\rm 2h}/d\ln k$ would imply that the profiles do change (in the sense
that they are fixed in comoving, rather than physical coordinates).

We now turn to the one-halo term. Given our assumption about density profiles,
this term is much simpler. The only effect is the change in the mass function,
which through \refeq{bNdef} (for $N=1$) yields
\be
\frac{\partial}{\partial\bar\d} I^0_2(k, k) = I^1_2(k,k)\,.
\ee
We thus obtain
\ba
\frac{d P^{\rm 1h}(k)}{d\bar\d}\Big|_0 =\:& I^1_2(k,k)\,.
\label{eq:P1hd0}
\ea
Putting everything together, we obtain
\be
\frac{d\ln P_{\rm HM}(k)}{d\bar\d}\Big|_0 = \left[P_{\rm HM}(k)\right]^{-1} 
\left[ \left(\frac{68}{21} - \frac13 \frac{d\ln k^3 P_l(k)}{d\ln k}\right) P^{\rm 2h}(k)
 + 2 I^2_1(k) I^1_1(k) P_l(k) + I^1_2(k,k) \right]\,.
\label{eq:Pkd0HM}
\ee
The prediction of \refeq{Pkd0HM} is shown as the blue lines in
\reffig{sep_uni_nonpert}. The amplitude and broad-band shape agree with
the simulations well. The main discrepancy in the halo model prediction
is the insufficient damping of the BAO wiggles. 

An alternative approach to derive the halo model prediction for $iB(k)$ is to
use higher $N$-point functions
\cite{kehagias/perrier/riotto:2013,li/hu/takada:2014}, which are
decomposed into $1-,\dots, N-$halo terms. We now compare \refeq{Pkd0HM}
with the results of \cite{li/hu/takada:2014}, which were derived from
the halo model four-point function in the collapsed limit. Note that the
squeezed limit is assumed in both approaches. Their eq.~(27) is
\be
\frac{d\ln P_{\rm HM}(k)}{d\bar\d}\Big|_0 = \left[P_{\rm HM}(k)\right]^{-1} 
\left[ \left(\frac{68}{21} - \frac13 \frac{d\ln k^3 P^{\rm 2h}(k)}{d\ln k}\right)P^{\rm 2h}(k) + I^1_2(k,k) \right]\,.
\label{eq:LHT}
\ee
There are two differences to \refeq{Pkd0HM}: the term $\propto I^2_1$ is
absent, and the slope is taken from from $P^{\rm 2h}$ rather than
$P_l$. The $I^2_1$ term is absent in \refeq{LHT} as by assumption $b_2$
was taken to be zero in the four-point function of
\cite{li/hu/takada:2014}; as discussed above, its inclusion is somewhat
ambiguous given the lack of self-consistency of the halo model
approach. The different power spectrum slopes are due to the different
sources of this term in the two derivations. In our case, the assumption
of unchanged halo profiles dictates the form of \refeq{Pkd0HM}. In the
derivation of \refeq{LHT}, the slope originates from the integral over
the $F_2$ kernel in the 3-halo term, which proceeds as described in
\refapp{tr_bi_sq} but involves $P^{\rm 2h}$ instead of $P_l$. Note
however that the numerical difference between \refeq{LHT} and
\refeq{Pkd0HM} is only at the percent level.

\subsection{Dependence on cosmological parameters}
\label{sec:cosmodep}

Both the matter power spectrum and (integrated) bispectrum depend on the
cosmological parameters such as $\Om,\,\sigma_8, n_s$. However, the normalized
integrated bispectrum is much less sensitive to cosmology as the leading
cosmology dependence is taken out by the normalizing denominator.

\refEq{Pkd01loop} is useful for understanding the dependence of the
response function of the power spectrum (and thus the normalized
integrated bispectrum) on cosmological parameters. The second term
depends on the local spectral index of the matter power spectrum,
$d\ln k^3P(k)/d\ln k$, which depends on the initial power spectrum tilt,
$n_s$, and the matter and radiation densities which change the redshift
of matter-radiation equality as well as the BAO scale.  
It also depends on the shape of BAO wiggles,
and increasing the amplitude of the matter power spectrum ($\sigma_8$)
leads to a stronger damping of the BAO feature.  Increasing $\sigma_8$ further 
increases the last term, which is proportional to $\sigma_8^2$.  
\begin{figure}[t]
\centering{
\includegraphics[width=1\textwidth]{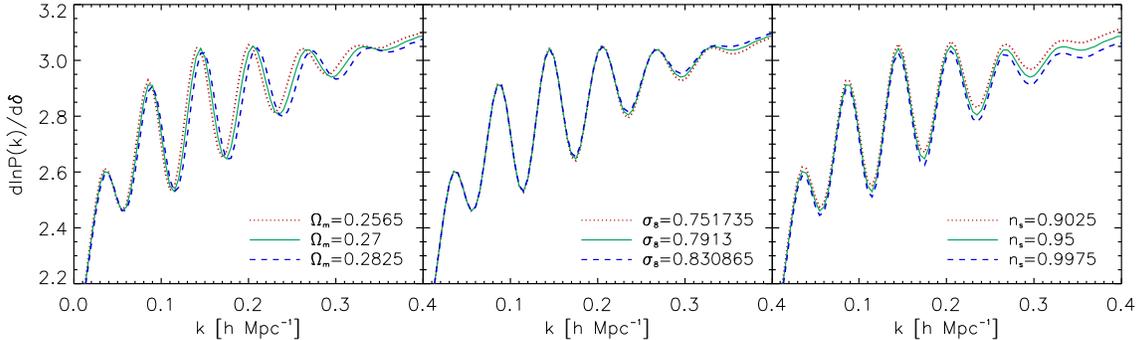}
}
\caption{The linear response functions computed from the SPT 1-loop power
 spectrum with various cosmological parameters at $z=2$. The fiducial cosmology
 ($\Omega_m=0.27$, $\sigma_8=0.7913$, and $n_s=0.95$) is shown in green
 solid lines. The red dotted and green dashed lines represent the cosmologies
 with $\pm 5\%$ of the fiducial parameters, $\Omega_m$ (left), $\sigma_8$
 (middle), and $n_s$ (right).
}
\label{fig:cosmo_dep}
\end{figure}
\refFig{cosmo_dep} shows the linear response functions, $d\ln P(k,a)/d\bar\d(a)$
computed from the SPT 1-loop power spectrum (\refeq{Pkd01loop}) at $z=2$
when varying cosmological parameters by $\pm 5\%$.  
The effects on the response functions are at the percent level or less,
illustrating the weak cosmology dependence of this observable.  On the
scales considered, the shift in the BAO scale when varying $\Om$ leads
to the relatively largest effect.  We expect that the sensitivity to changes
in $\sigma_8$ will be higher on smaller, more nonlinear scales.  

\section{Conclusions}
\label{sec:conclusion}
In this paper, we have proposed a novel method to measure the
squeezed-limit bispectrum. By correlating the mean density fluctuation
and the position-dependent power spectrum, we obtain a measurement
of a certain moment of the bispectrum (integrated bispectrum) without
having to actually measure three-point correlations in the data.  The
integrated bispectrum is dominated by the squeezed-limit bispectrum,
which is much easier to model than the full bispectrum for all configurations.  
This is evidenced by figures~\ref{fig:sep_uni_pert}--\ref{fig:sep_uni_nonpert},
where we show model predictions accurate to a few percent using existing techniques
and without tuning any parameters.

A further, key advantage of this new observable is that both the mean density
fluctuation and the power spectrum are significantly easier to measure
in actual surveys than the bispectrum in terms of survey selection
functions. In particular, the procedures developed for power spectrum
estimation can be directly applied to the measurement of the
position-dependent power spectrum. Additionally, the position-dependent
power spectrum depends on only one wavenumber (at fixed size of the
subvolume) rather than the three wavenumbers of the
bispectrum. Consequently, the covariance matrix also becomes easier to
model.

We have measured the position-dependent power spectrum in 160
collisionless $N$-body simulations with Gaussian initial conditions, and
have used two different approaches --- bispectrum modeling and the
separate universe approach --- to model the measurements. All of the
approaches work well on large scales, $k \lesssim 0.2~h\,{\rm Mpc}^{-1}$,
and at high redshift. On small scales, where non-linearities
become important, the separate universe approach (\refsec{sep_uni})
applied through the Coyote emulator prescription performs best at redshifts
$z < 2$, while the SPT 1-loop predictions perform equally well at $z\geq 2$.  
Both show agreement to within a few percent up to $k = 0.4~h\,{\rm Mpc}^{-1}$.
Accurate predictions for the position-dependent power spectrum on these
and even smaller scales can be obtained by applying the separate universe 
approach to  dedicated small-box $N$-body simulations of curved 
cosmologies \cite{li/hu/takada:2014}.  We shall study this in an upcoming paper.

The normalized integrated bispectrum is relatively insensitive to changes in
cosmological parameters (\refsec{cosmodep}), and we do not expect that it will
allow for competitive cosmology constraints.  On the other hand, this property
can also be an advantage:  since this observable can be predicted accurately
without requiring a precise knowledge of the cosmology, it can serve as a
useful systematics test for example in weak lensing surveys.  As an example,
consider \refeq{iB_sq} applied to shear measurements.  A constant 
multiplicative bias $1+m$ in the shear estimation contributes a factor 
$(1+m)^3$ on the left hand side of the equation, and a factor $(1+m)^4$ on 
the right hand side.  Thus, by comparing the measured normalized integrated
bispectrum with the (essentially cosmology-independent) expectation, one
can constrain the multiplicative shear bias.

The position-dependent power spectrum can also naturally be applied to the case of
spectroscopic galaxy surveys, in which case the non-linear bias of the observed tracers 
also contributes to the bispectrum and position-dependent power spectrum.  
Thus, when applied to halos or galaxies, this observable
can serve as an independent probe of the bias parameters and break degeneracies
between bias and growth which are present when only considering the halo or galaxy power
spectrum. We shall apply this new method to halos in $N$-body simulations, as
well as to data from galaxy surveys in future papers.  
Finally, this approach can also be immediately applied to the projected
matter density distribution as measured through weak lensing.  In this case,
the complexities of bias are absent and the modeling we have presented in this
paper should be sufficient to describe the measurements.

\acknowledgments
We would like to thank Wayne Hu, Masahiro Takada, Simon White,
and Donghui Jeong for useful discussions.

\appendix
\section{Tree-level matter bispectrum in the squeezed configurations}
\label{app:tr_bi_sq}

In this appendix, we derive the squeezed-limit result \refeq{bi_approx}.  
The tree-level perturbation theory gives the matter bispectrum (with our
notation) 
\begin{eqnarray}
 B(\vk-\vq_1,-\vk+\vq_1+\vq_3,-\vq_3)&=&
 2[F_2(\vk-\vq_1,-\vk+\vq_1+\vq_3)
 P(|\vk-\vq_1|)P(|-\vk+\vq_1+\vq_3|) \nonumber\\
 &&+F_2(\vk-\vq_1,-\vq_3)P(|\vk-\vq_1|)P(q_3) \nonumber\\
 &&+F_2(-\vk+\vq_1+\vq_3,-\vq_3)
 P(|-\vk+\vq_1+\vq_3|)P(q_3)] ~,
\end{eqnarray}
where $F_2(\vk_1,\vk_2)$ is
\begin{equation}
 F_2(\vk_1,\vk_2)=\frac{5}{7}+\frac{1}{2}\frac{\vk_1\cdot\vk_2}{k_1k_2}
 \left(\frac{k_1}{k_2}+\frac{k_2}{k_1}\right)+\frac{2}{7}
 \left(\frac{\vk_1\cdot\vk_2}{k_1k_2}\right)^2 ~.
\end{equation}
In the squeezed configurations, where $k\gg q_1,q_3$, we Taylor expand
the power spectra and $F_2$'s. In the calculation, we keep terms to
first order, e.g., keeping 1 and $q/k$ (ignoring $(q/k)^n$ for $n\ge2$)
or keeping $q/k$ and $(q/k)^2$ (ignoring $(q/k)^n$ for $n\ge3$), and
then combine them to see the leading order effect of the final result. 

First, the amplitudes of the vectors can be calculated as
\begin{eqnarray}
 |\vk-\vq_1|&=&\sqrt{k^2+q_1^2-2\vk\cdot\vq_1}
 \approx k\left(1-\frac{\vk\cdot\vq_1}{k^2}\right) \nonumber\\
 |-\vk+\vq_1+\vq_3|&\approx&k
 \left[1-\frac{\vk\cdot(\vq_1+\vq_3)}{k^2}\right] ~.
\end{eqnarray}
Therefore, the power spectra are
\begin{eqnarray}
 P(|\vk-\vq_1|)&\approx&
 P\left(k-\frac{\vk\cdot\vq_1}{k}\right)
 \approx P(k)-\frac{\vk\cdot\vq_1}{k}\frac{dP(k)}{dk}
 =P(k)\left[1-\frac{\vk\cdot\vq_1}{k^2}\frac{d\ln P(k)}{d\ln k}\right] \nonumber\\
 P(|-\vk+\vq_1+\vq_3|)&\approx&
 P(k)\left[1-\frac{\vk\cdot(\vq_1+\vq_3)}{k^2}\frac{d\ln P(k)}{d\ln k}\right] ~.
\end{eqnarray}

The cosines between the vectors forming the squeezed triangle are
\begin{eqnarray}
 \frac{(\vk-\vq_1)\cdot(-\vk+\vq_1+\vq_3)}
 {|\vk-\vq_1||-\vk+\vq_1+\vq_3|}&\approx&
 [-k^2+\vk\cdot(2\vq_1+\vq_3)]\frac{1}{k^2}
 \left[1+\frac{\vk\cdot\vq_1}{k^2}\right]
 \left[1+\frac{\vk\cdot(\vq_1+\vq_3)}{k^2}\right]\approx-1 \nonumber\\
 \frac{(\vk-\vq_1)\cdot(-\vq_3)}{|\vk-\vq_1||-\vq_3|}
 &\approx&\frac{1}{kq_3}\left[-\vk\cdot\vq_3
 -\frac{(\vk\cdot\vq_1)(\vk\cdot\vq_3)}{k^2}
 +\vq_1\cdot\vq_3\right] \nonumber\\
 \frac{(-\vk+\vq_1+\vq_3)\cdot{-\vq_3}}
 {|-\vk+\vq_1+\vq_3||-\vq_3|}&\approx&
 \frac{1}{kq_3}\left[\vk\cdot\vq_3
 +\frac{(\vk\cdot\vq_3)[(\vk\cdot(\vq_1+\vq_3)]}{k^2}
 -\vq_3\cdot(\vq_1+\vq_3)\right] ~,
\end{eqnarray}
the terms $k_1/k_2+k_2/k_1$ are
\begin{eqnarray}
 \frac{|\vk-\vq_1|}{|-\vk+\vq_1+\vq_3|}
 +\frac{|-\vk+\vq_1+\vq_3|}{|\vk-\vq_1|}
 &\approx&2 \nonumber\\
 \frac{|\vk-\vq_1|}{|-\vq_3|}+\frac{|-\vq_3|}{|\vk-\vq_1|}
 &\approx&\frac{k}{q_3}\left[1-\frac{\vk\cdot\vq_1}{k^2}\right] \nonumber\\
 \frac{|-\vk+\vq_1+\vq_3|}{|-\vq_3|}
 +\frac{|-\vq_3|}{|-\vk+\vq_1+\vq_3|}&\approx&
 \frac{k}{q_3}\left[1-\frac{\vk\cdot(\vq_1+\vq_3)}{k^2}\right] ~,
\end{eqnarray}
and thus the $F_2$'s become
\begin{eqnarray}
 F_2(\vk-\vq_1,-\vk+\vq_1+\vq_3)&\approx&0 \nonumber\\
 F_2(\vk-\vq_1,-\vq_3)&\approx&\frac{5}{7}
 +\frac{1}{14(kq_3)^2}[-7k^2(\vk\cdot\vq_3)+7k^2(\vq_1\cdot\vq_3)
 +4(\vk\cdot\vq_3)^2] \nonumber\\
 F_2(-\vk+\vq_1+\vq_3,-\vq_3)&\approx&
 \frac{5}{7}+\frac{1}{14(kq_3)^2}[7k^2(\vk\cdot\vq_3)
 -7k^2[\vq_3\cdot(\vq_1+\vq_3)]+4(\vk\cdot\vq_3)^2] ~. \nonumber\\
\end{eqnarray}
Finally, we combine all terms, keep the leading order terms, and obtain
\begin{eqnarray}
 B(\vk-\vq_1,-\vk+\vq_1+\vq_3,-\vq_3) &=&
 \left[\frac{13}{7}+\frac{8}{7}\left(\frac{\vk\cdot\vq_3}{kq_3}\right)^2
 -\left(\frac{\vk\cdot\vq_3}{kq_3}\right)^2\frac{d\ln P(k)}{d\ln k}\right]
 P(k)P(q_3) \vs
& & + \cO\left(\frac{q_{1,3}}{k}\right) ~,
\end{eqnarray}
where $q_{1,3}$ refers to $\vq_1$ and $\vq_3$. Spherically averaging
$(\hat\vk\cdot\hat\vq_3)^2$ over $\vk$ yields 1/3, and thus
\begin{eqnarray}
 \int\frac{d^2\Omega_{\hat\vk}}{4\pi}~
 B(\vk-\vq_1,-\vk+\vq_1+\vq_3,-\vq_3)&=&
 \left[\frac{47}{21}-\frac{1}{3}\frac{d\ln P(k)}{d\ln k}\right]P(k)P(q_3)
 + \cO\left(\frac{q_{1,3}}{k}\right)^2 \nonumber\\
 &=&\left[\frac{68}{21}-\frac{1}{3}\frac{d\ln k^3 P(k)}{d\ln k}\right]P(k)P(q_3)
 + \cO\left(\frac{q_{1,3}}{k}\right)^2 ~. \nonumber\\
\end{eqnarray}
Note that the $\cO(q_{1,3}/k)$ terms cancel in the angular average.
The same relation has been derived in
\cite{valageas:2013,li/hu/takada:2014,kehagias/perrier/riotto:2013,figueroa/etal:2012}.

\bibliography{references}
\end{document}